\documentclass[
preprint,
]{revtex4-2}
\usepackage[dvipdfmx]{hyperref}
\usepackage{graphicx}% Include figure files
\usepackage{dcolumn}% Align table columns on decimal point
\usepackage{bm}% bold math
\usepackage{indentfirst}
\usepackage{here}
\usepackage{comment}
\usepackage{hyperref}% add hypertext capabilitie
\hypersetup{colorlinks=true,linkcolor=blue,citecolor=blue,filecolor=blue,urlcolor=blue}
\begin{document}

% \preprint{APS/123-QED}

\title{A high-temperature multiferroic Tb$_2$(MoO$_4$)$_3$}

\author{Shimon Tajima}
\email{shimon.tajima.p2@dc.tohoku.ac.jp}
\author{Hidetoshi Masuda}
\affiliation{Institute for Materials Research, Tohoku University, Sendai 980-8577, Japan}
\author{Yoichi Nii}
\affiliation{Institute for Materials Research, Tohoku University, Sendai 980-8577, Japan}
\author{Shojiro Kimura}
\affiliation{Institute for Materials Research, Tohoku University, Sendai 980-8577, Japan}
\author{Yoshinori Onose}
\affiliation{Institute for Materials Research, Tohoku University, Sendai 980-8577, Japan}

\begin{abstract}
\bf{Magnetoelectric mutual control in multiferroics, which is the electric control of magnetization, or reciprocally the magnetic control of polarization has attracted much attention because of its possible applications to spintronic devices, multi-bit memories, and so on.
While the required working temperature for the practical application is much higher than room temperature, which ensures stable functionality at room temperature, the reported working temperatures were at most around room temperature. Here, we demonstrated magnetic control of ferroelectric polarization at 432 K in ferroelectric and ferroelastic Tb$_2$(MoO$_4$)$_3$, in which the polarity of ferroelectric polarization is coupled to the orthorhombic strain below the transition temperature 432 K. The paramagnetic but strongly magnetoelastic Tb$^{3+}$ magnetic moments enable the magnetic control of ferroelectric and ferroelastic domains; the ferroelectric polarization is controlled depending on whether the magnetic field is applied along [110] or [$\bar{1}$10]. This result may pave a new avenue for designing high-temperature multiferroics.}
\end{abstract}

\maketitle
The highest working temperature of multiferroics is at present around room temperature\cite{review_RT}-\cite{Y-hexaferrite_cross-control}. To achieve higher working temperature that enables commercial application\cite{4state_memory}-\cite{MF_review2022}, a new material strategy should be needed. Multiferroics can be classified into two categories\cite{ME_memory}, \cite{Class_multiferro},\cite{BFO_review}.
In one category, ferroelectricity is induced by some magnetic ordering. In this case, the coupling between polarization and magnetism is very large, and a giant magnetoelectric effect is observed. Nevertheless, the ferroelectric transition temperature is usually lower than room temperature\cite{TbMnO3}-\cite{hexaferrite_M-reversal}. The exceptional case is hexaferrite\cite{Z-hexaferrite_RT}-\cite{Y-hexaferrite_cross-control},\cite{hexaferrite_science},\cite{hexaferrite_M-reversal}. The ferroelectric transition temperature is reported to be around 450 K but the working temperature of the magnetoelectric effect is at most around room temperature due to the problem of residual electric conduction. In the other category, ferroelectricity emerges independently of magnetic ordering. Several multiferroic materials in this category have ferroelectric transition temperatures much higher than room temperature\cite{ME_memory},\cite{Class_multiferro},\cite{BFO_review}. In this case, the coupling between electric polarization and magnetism is usually very weak so that the ferroelectric polarization can not be reversed by a magnetic field. In this paper, we demonstrated that the magnetoelectric coupling becomes strong, and the high-temperature magnetoelectric effect is achieved in a latter type of multiferroic Tb$_2$(MoO$_4$)$_3$, which is ferroelastic as well as ferroelectric and has paramagnetic Tb$^{3+}$ moments with strong magnetoelastic coupling.

\begin{figure}[h]
    \centering
    \includegraphics[width=8cm]{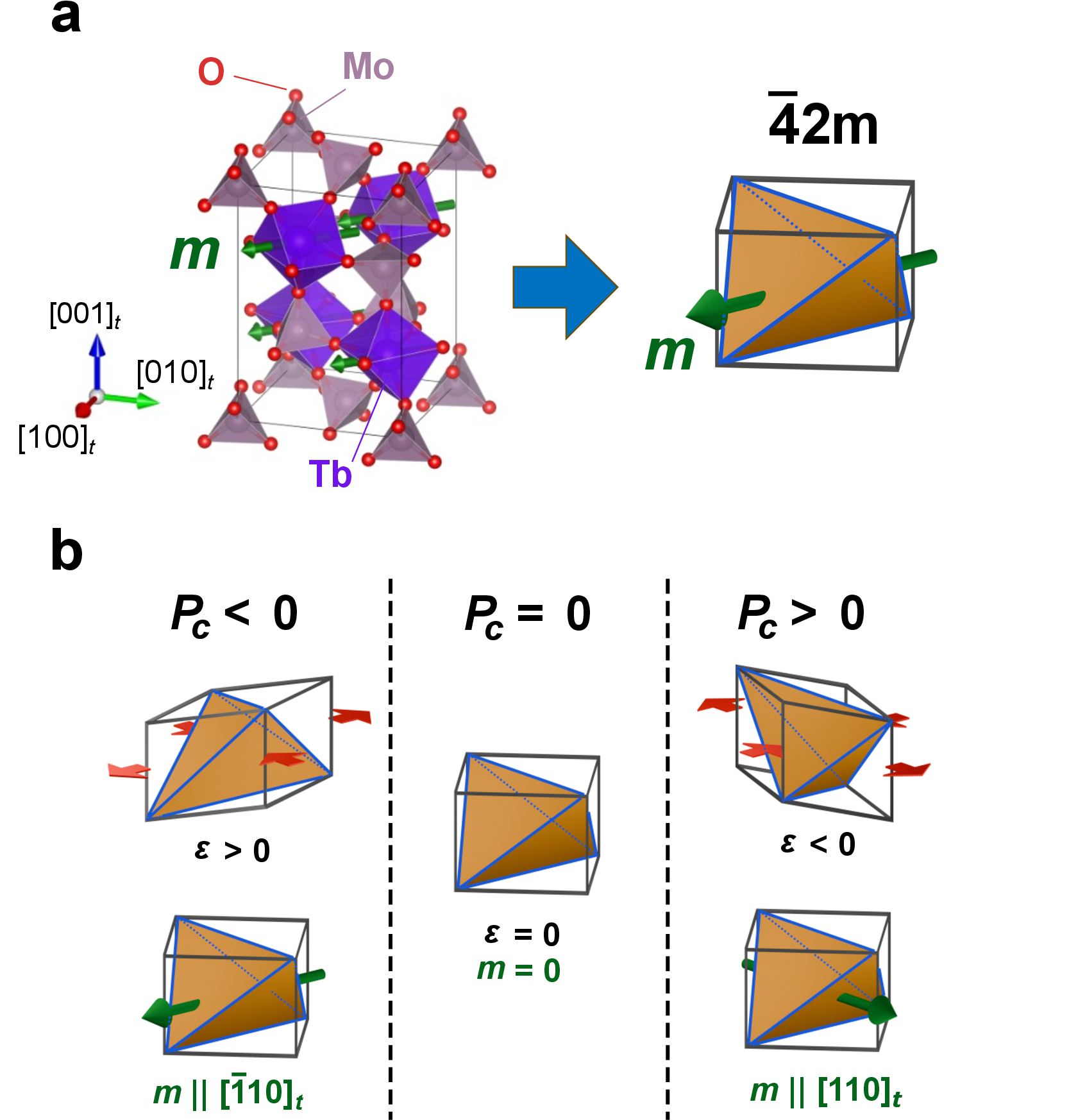}
    \caption{Piezoelectric and magnetoelectric properties symmetrically expected in Tb$_2$(MoO$_4$)$_3$. (\textbf{a}) Schematic illustrations of tetragonal crystal structure of Tb$_2$(MoO$_4$)$_3$ above $T_c$ = 432 K, and a tetrahedron, which is a motif of $\bar{4}$2m point group symmetry, with a magnetic moment. The crystal structure is drawn by VESTA\cite{vesta}. (\textbf{b}) Schematic illustrations of piezoelectric and magnetoelectric properties of $\bar{4}$2m point group. Uniaxial stress $\varepsilon$ along the [110]$_t$ or [$\bar{1}1$0]$_t$ direction induces the polarization. The sign of polarization depends on whether the stress is applied along [110]$_t$ ($\varepsilon>0$) or [$\bar{1}1$0]$_t$ ($\varepsilon<0$). Similarly, the polarization can be induced by a magnetic moment and the sign depends on whether it is along [110]$_t$ or [$\bar{1}1$0]$_t$.}
    \label{TMO_structure}
\end{figure}

Figure \ref{TMO_structure}{\bf a} shows the crystal structure of Tb$_2$(MoO$_4$)$_3$ above the ferroelectric transition temperature $T_c$ = 432 K\cite{TMO_struc_HT}. It is composed of corner-shared MoO$_4$ tetrahedra and TbO$_7$ octahedra. The space group of this high-temperature paraelectric state is $P\bar{4}2_1$m and the point group is $\bar{4}2$m, which is not polar but noncentrosymmetric.
Several multiferroic materials such as CuB$_2$O$_4$ and Ba$_2$CoGe$_2$O$_7$ have the same point group and show unique magnetoelectric properties reflecting the point group symmetry\cite{Ba2CoGe2O7_ME},\cite{CuB2O4_optical},\cite{Ba2CoGe2O7_optical}. To illustrate the symmetrical proprieties of $\bar{4}2$m point group, we utilize a compressed tetrahedron. It is a simple object and has the $\bar{4}2$m point group symmetry as shown in Fig. \ref{TMO_structure}{\bf a}. The electric polarization along the [001]$_t$ direction should be piezoelectrically induced by a diagonal uniaxial strain in the (001)$_t$ plane (Fig. \ref{TMO_structure}{\bf b}). Here, the suffix {\it t} denotes the crystal axes in the notation of high temperature tetragonal phase. The sign of electric polarization depends on whether the uniaxial strain is along [110]$_t$ or [$\bar{1}$10]$_t$. Importantly, electric polarization is similarly induced by a magnetic moment. When the magnetic moment is along [110]$_t$, the up-down symmetry is broken and polarization should be induced. The polarization is unchanged by the inversion of the magnetic moment but reversed by the 90 $^\circ$ rotation of the magnetic moment. Because of this symmetrical property, the strong coupling between the ferroelectric polarization and magnetic moments is expected in this material. In this paper, we have certainly demonstrated the magnetic control of electric polarization below $T_c$ = 432 K.

% \noindent\textbf{Results}\par
\section*{Results}
\begin{figure}[h]
    \centering
    \includegraphics[width=8cm]{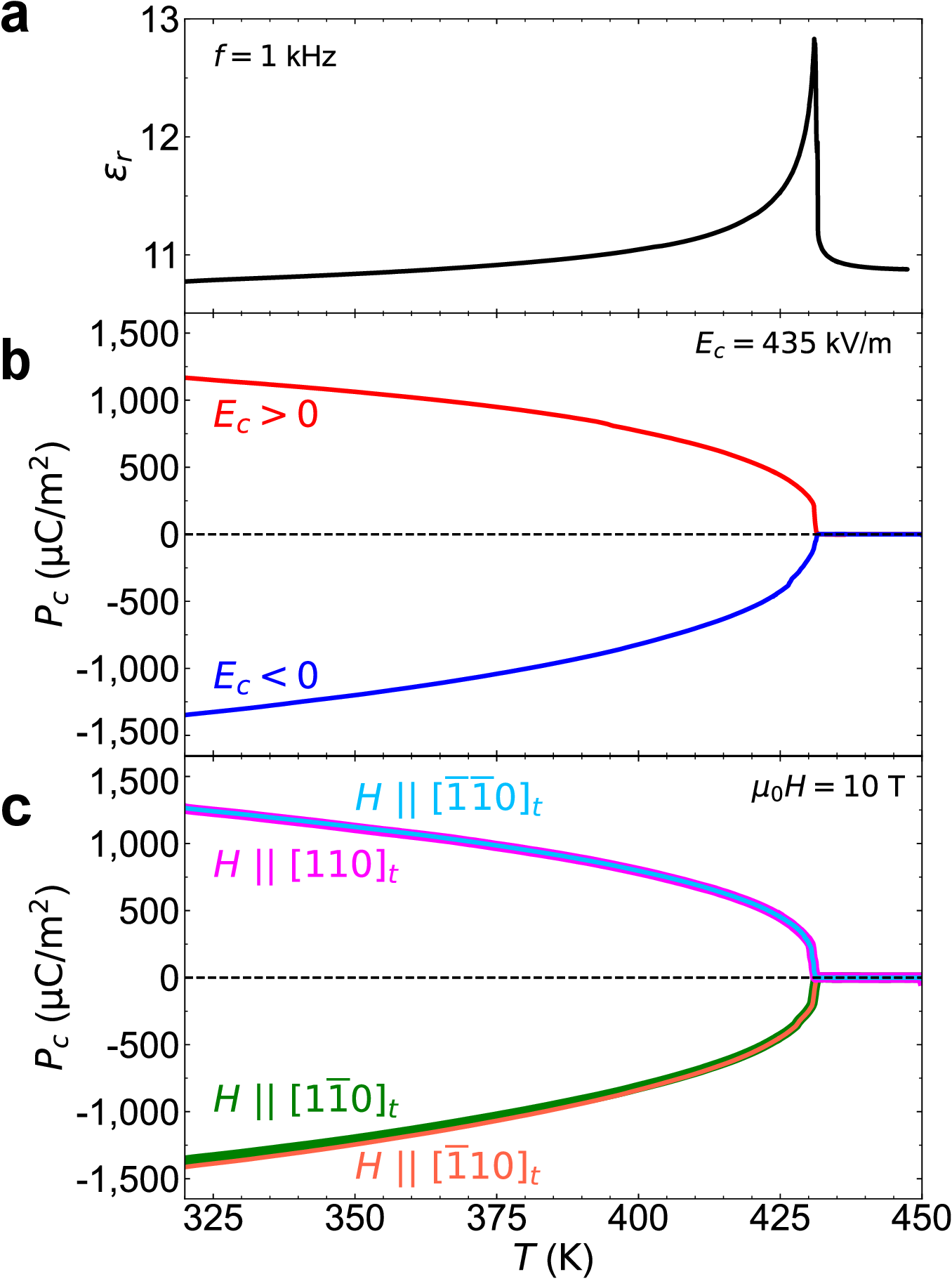}
    \caption{Magnetically induced polarization in magnetic fields. (\textbf{a}) Temperature dependence of dielectric constant at $\mu_0H=0$ T. (\textbf{b}) Temperature dependence of the electric polarization at $\mu_0H=0$ T. Before the measurement, we cooled the sample in electric fields $E_c=\pm435$ kV/m. (\textbf{c}) Temperature dependence of the electric polarization measured on warming runs in the absence of any external field after cooling the sample down to 320 K applying magnetic fields $|\mu_0H|=10$ T with various directions.}
    \label{TMO_basic_properties}
\end{figure}

Figures \ref{TMO_basic_properties}{\bf a, b} show the temperature ($T$) dependences of relative dielectric constant ($\varepsilon_r$) and spontaneous polarization along the $c$-axis around $T_c$ = 432 K for Tb$_2$(MoO$_4$)$_3$. Before the spontaneous polarization measurement, we applied an electric field $E_c$ = $\pm$435 kV/m, cooled the sample from 450 K to 320 K, and then turned off the electric field. Finally, we measured the polarization with increasing temperature. The dielectric constant was also measured with increasing temperature. The relative dielectric constant shows a sharp peak at $T_c$ = 432 K and the spontaneous polarization depending on the signs of $E_c$ emerges below $T_c$, being consistent with the literature\cite{RMO_FE},\cite{TMO_FE}. As reported previously, the ferroelectric transition is simultaneously a ferroelastic transition\cite{TMO_FE}. The two diagonal lengths in the $c$ plane of the crystal structure become different from each other, and the orthorhombic distortion is coupled to the ferroelectric polarization below $T_c$. Therefore, this transition can be viewed as the spontaneous induction of uniaxial strain along [110]$_t$ or [$\bar{1}$10]$_t$ shown in Fig. \ref{TMO_structure}{\bf b}.

One can expect that the ferroelectric/ferroelastic domain can be controlled by a magnetic field according to the inference based on Fig.\ref{TMO_structure}{\bf b}. We certainly demonstrated that the ferroelectric polarization can be controlled by a magnetic field. Figure \ref{TMO_basic_properties}{\bf c} shows the temperature dependence of ferroelectric polarization measured on warming run in the absence of any electric and magnetic fields after cooling in magnetic fields with magnitudes of 10 T and various directions. The magnitude of polarization is comparable with that after cooling in an electric field, which indicates that the magnetic field effectively aligns the ferroelectric domain. More importantly, the sign of polarization depends on whether the magnetic field applied before the polarization measurement is along [110]$_t$ or [$\bar{1}$10]$_t$ while it is unchanged by the reversal of the magnetic field. It should be noted that similar magnetoelectric properties were observed in the multiferroics with the same $\bar{4}$2m point group symmetry such as Ba$_2$CoGe$_2$O$_7$\cite{Ba2CoGe2O7_ME}. While the magnetoelectric responses were measured in the antiferromagnetic state in the previous cases, the present observation was in the paramagnetic state below the high transition temperature of 432 K\cite{TMO_FE},\cite{TMO_AFM}.

\begin{figure}[h]
    \centering
    \includegraphics[width=8cm]{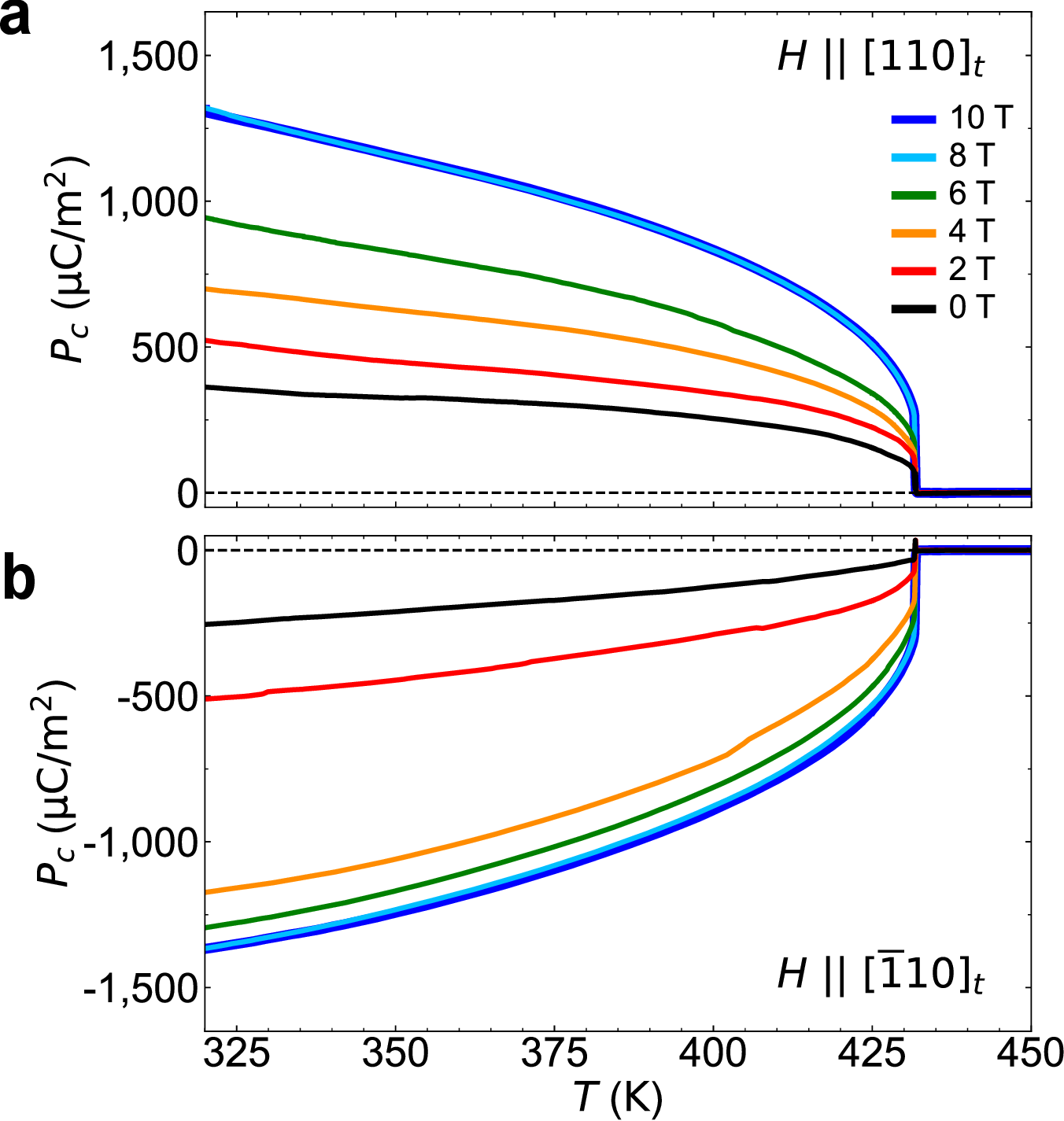}
    \caption{Temperature dependence of the electric polarizations measured on cooling runs in various magnetic fields along (\textbf{a}) [110]$_t$ and (\textbf{b}) [$\bar{1}1$0]$_t$.}
    \label{TMO_P-T_H}
\end{figure}

Figures \ref{TMO_P-T_H}{\bf a, b} show the temperature dependence of electric polarization measured on cooling runs in various magnetic fields along [110]$_t$ and [$\bar{1}$10]$_t$, respectively. Even at zero magnetic fields, finite electric polarization was observed, which indicates that the positive and negative ferroelectric domains were not perfectly canceled. In fact, the sign and magnitude of polarization at zero magnetic fields were not reproducible so the 0 T data in Fig. \ref{TMO_P-T_H}{\bf a} is different from that in Fig. \ref{TMO_P-T_H}{\bf b}. The positive electric polarization increases with increasing the magnetic field along [110]$_t$. Above 8 T, the electric polarization was almost saturated. When the magnetic field was applied along [$\bar{1}$10]$_t$, the polarization was negative but the magnitude showed similar magnetic field dependence.

\begin{figure}[h]
    \centering
    \includegraphics[width=13cm]{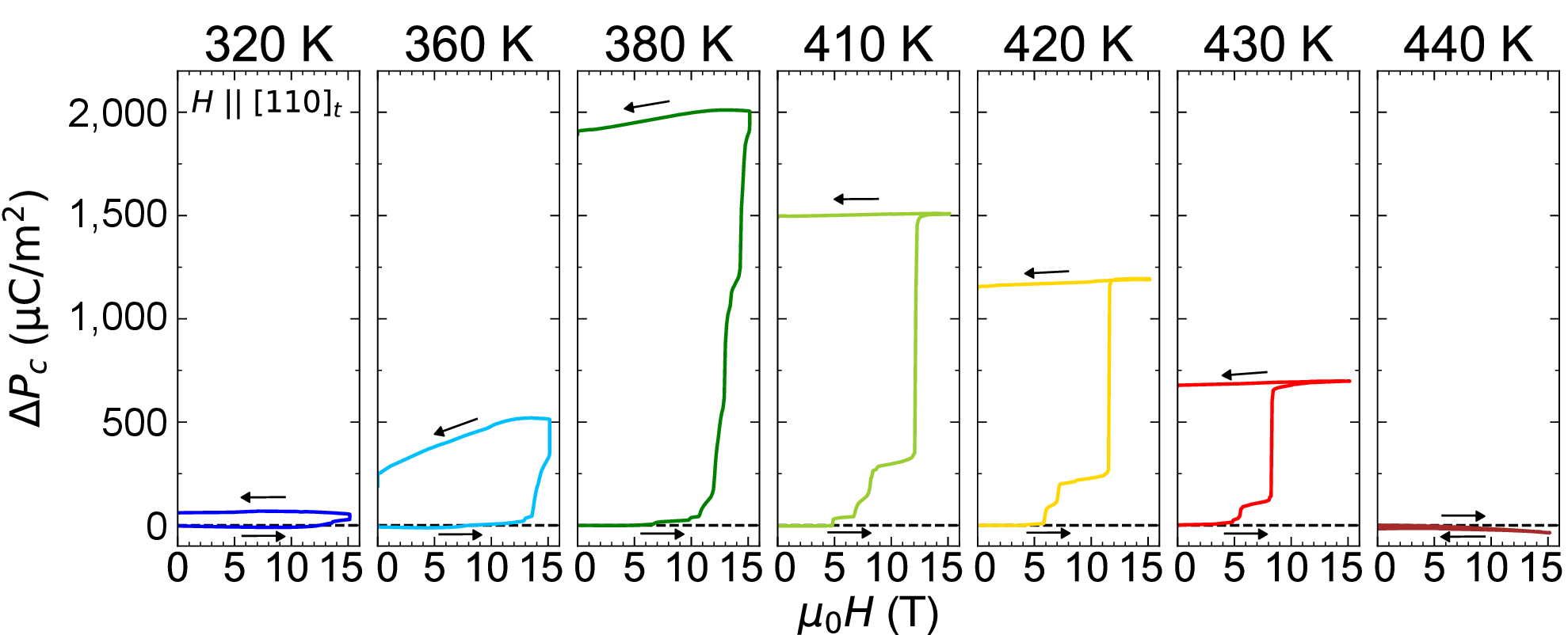}
    \caption{Magnetic reversal of electric polarization. Before the measurements, the polarization is aligned along the negative direction by an electric field. Then we applied magnetic fields along [110]$_t$ up to 15 T and then decreased it to 0 T at various temperatures while measuring the polarization.}
    \label{TMO_dP-H}
\end{figure}

Finally, we demonstrated the magnetic reversal of ferroelectric polarization. Figure \ref{TMO_dP-H} shows the magnetic field variation of electric polarization at various temperatures. Before the measurements, we aligned the electric polarization along the negative direction with an electric field. Then, we increased the magnetic field along [110]$_t$ to 15 T and decreased it to 0 T while measuring the electric polarization. Note that the magnetic field along [110]$_t$ favors the positive polarization state and therefore the polarization reversal is expected if the magnetic field is strong enough. At 430 K, we observe an abrupt change of polarization around 9 T which corresponds to the polarization reversal. While a similar polarization reversal was previously observed below 100 K\cite{TMO_P-reverse},\cite{RMO_ME}, we here demonstrate it is possible at this high temperature. As the temperature decreases from 430 K, the reversal magnetic field and the magnitude of polarization change increase. The increase of polarization magnitude is consistent with the evolution of ferroelectric polarization toward low temperature shown in Fig. \ref{TMO_P-T_H}. Below 360 K, the polarization reversal was not fully observed because the reversal magnetic field is larger than 15 T. Previously, the temperature of magnetic polarization reversal was at most around room temperature\cite{Y-hexaferrite_RT},\cite{Y-hexaferrite_cross-control}. To our knowledge, this is the highest temperature of magnetic reversal of ferroelectric polarization.

\section*{Discussion}
In summary, we have successfully demonstrated magnetic control of ferroelectric polarization at 430 K. While the spin-dependent metal-ligand hybridization mechanism was proposed as a mechanism of related multiferroic compounds such as Ba$_2$CoGe$_2$O$_7$\cite{Ba2CoGe2O7_ME},\cite{d-p_hybridization}, the expected small hybridization to the localized Tb 4f state does not support it as the mechanism of magnetoelectric response in the present material. Because Tb$^{3+}$ ion is known for its strong magnetoelastic coupling\cite{TMO_P-reverse}, the combination of piezoelectricity and magnetoelastic coupling is most likely the mechanism of magnetoelectric response in this system. The piezoelectricity couples the ferroelectric polarization to the orthorhombic distortion, while the magnetoelastic coupling couples the orthorhombic distortion to magnetic anisotropy. Thus, these two effects result in the coupling between magnetization direction and polarization. The point group symmetry $\bar{4}$2m enables the novel connection between the ferroelectric polarization and magnetic moments. While several multiferroic mechanisms have been proposed, this type of mechanism has scarcely been discussed as a multiferroic mechanism\cite{TMO_P-reverse},\cite{RMO_ME}. In this sense, this work may pave a new avenue for exploring high-temperature multiferroics.

\section*{Methods}
We grew single crystals of Tb$_2$(MoO$_4$)$_3$ in the air utilizing the floating-zone method (Supplementary Fig. 1 and Supplementary Fig. 2)\cite{TMO_FZ}. For the measurements of dielectric constant and electric polarization, we used a rectangular piece of single crystal with the size of $0.73\times1.28\times0.46$ mm$^3$. The widest surface is parallel to the $c$ plane. Before the measurements, we annealed the sample at 1000 $^{\circ}$C in the air to remove possible stress in the sample (Supplementary Fig. 3). We measured the polarization and dielectric constant in a cryogen free superconducting magnet in the High Field Laboratory for Superconducting Materials, Institute for Materials Research, Tohoku University utilizing an electrometer (KEITHLEY 6517A) and a capacitance bridge (Andeen-Hagerling AH2700A). For the measurements of electric polarization, we have calibrated the effect of a small background current ($\sim$ 0.1 pA).

\section*{Acknowledgements}
The measurements were carried out at the High Field Laboratory for Superconducting Materials, IMR, Tohoku University. We thank Prof. Fujita,  Prof. Nojima, and Prof. Umetsu for their help of X-ray diffraction measurements. This work was supported by JSPS KAKENHI(Grants No. JP20K03828, No. JP21H01036, No. JP22H04461, No. JP23K13654, No. JP24H01638, No. JP24H00189, No. 21H01026, No. 23H04863, and No. 23K17660), JST SPRING(Grant No. JPMJSP2114), and JST PRESTO (Grant No. JPMJPR19L6). S. T. acknowledges support from GP-Spin at Tohoku University.

\end{document}